\documentclass[aps,prl,twocolumn,superscriptaddress,showpacs]{revtex4}

\usepackage{amsmath}
\usepackage{bm}
\usepackage{graphicx}

\begin{document}

\date{\today}

\title{Anomalous optical absorption in a random system with scale-free disorder}

\author{E.\ D\'{\i}az}
\affiliation{Departamento de F\'{\i}sica de Materiales, Universidad Complutense,
E-28040 Madrid, Spain}

\author{A.\ Rodr\'{\i}guez}
\thanks{Also at Grupo Interdisciplinar de Sistemas Complejos.}
\affiliation{Departamento de Matem\'{a}tica Aplicada y Estad\'{\i}stica,
Universidad Polit\'{e}cnica, E-28040 Madrid, Spain}

\author{F.\ Dom\'{\i}nguez-Adame}
\thanks{Also at Grupo Interdisciplinar de Sistemas Complejos.}
\affiliation{Departamento de F\'{\i}sica de Materiales, Universidad Complutense,
E-28040 Madrid, Spain}

\author{V.\ A.\ Malyshev}
\thanks{On leave from "S.I. Vavilov State Optical Institute", Birzhevaya Linia
12, 199034 Saint-Petersburg, Russia.}
\affiliation{Institute for Theoretical Physics and Materials Science Center,
University of Groningen, Nijenborgh 4, 9747 AG Groningen, The Netherlands}

\begin{abstract}

We report on an anomalous behavior of the absorption spectrum in a
one-dimensional lattice with long-range-correlated diagonal disorder with a
power-like spectrum in the form $S(k) \sim 1/k^{\alpha}$. These type of
correlations give rise to a phase of extended states at the band center,
provided $\alpha$ is larger than a critical value $\alpha_c$. We show that for
$\alpha < \alpha_c$ the absorption spectrum is single-peaked, while an
additional peak arises when $\alpha > \alpha_c$, signalling the occurrence of
the Anderson transition. The peak is located slightly below the low-energy
mobility edge, providing a unique spectroscopic tool to monitor the latter. We
present qualitative arguments explaining this anomaly.

\end{abstract}


\maketitle

Quantum dynamics of quasiparticles in random media has been a subject of
extensive studies since the seminal paper by Anderson, who argued that
quasiparticle states become localized for sufficiently large disorder, thus
giving rise to a localization-delocalization transition (LDT) in three
dimensions (3D)~\cite{Anderson58}. The hypothesis of single-parameter scaling,
introduced in Ref.~\onlinecite{Abrahams79}, lead to the general belief that all
eigenstates of noninteracting quasiparticles were localized  in one-~(1D) and
two dimensions, and that the LDT does not exist in low dimensional systems (for
recent overviews see Refs.~\onlinecite{Beenakker97} and~\onlinecite{Janssen98}).
However, at the end of the eighties and beginning of the nineties several
theoretical works provided clear evidences that short-range correlations in
disorder may cause delocalization even in 1D
systems~\cite{Flores89,Phillips91,Flores93,Adame93}. This fact  was put forward
to explain the high conductivity of doped polyaniline~\cite{Phillips91} as well
as the transport properties of GaAs-Al$_{x}$Ga$_{1-x}$As superlattices with
intentional correlated disorder~\cite{Bellani99}.

At the end of the last decade, it was demonstrated that long-range correlations
in disorder, with no characteristic spatial length (scale-free disorder), also
acts towards delocalization of 1D quasiparticle states~\cite{Moura98}. Random
sequences of correlated site energies characterized by a power-like spectrum
$S(k) \sim 1/k^{\alpha}$ with $\alpha > 0$, result in extended states provided
$\alpha$ is larger than some critical value $\alpha_c$~\cite{Moura98}. The
extended states form a phase at the band center, which is separated from
localized states by two mobility edges. This theoretical prediction was
experimentally validated by measuring microwave transmission spectra of a
single-mode waveguide with inserted correlated scatterers~\cite{Kuhl00}. The
observed transmission spectra were nicely simulated by the theoretical
model~\cite{Izrailev99}, confirming the existence of a phase of delocalized
states, in spite of the underlying randomness. Peculiarities of this type of
disorder had also their trace in biophysics, explaining the long-distance charge
transport in DNA sequences~\cite{Carpena02,Yamada04} as well as the existence of
a new class of level statistics~\cite{Carpena04}.

Recently, we have shown that 1D disordered systems with the above mentioned
correlated disorder in the site energies support Bloch-like
oscillations~\cite{Adame03}. The amplitude of the oscillations turned out to
carry information about the energy difference between the two mobility edges.
This finding opens the possibility to perform experiments on coherent dc charge
transport for measuring the bandwidth of the delocalized phase in disordered
systems with long-range correlated randomness. In this work we report on an
anomaly of the linear absorption within the underlined model. We show a
crossover of the absorption spectrum from a single-peaked to double-peaked shape
when the exponent $\alpha$ crosses the critical value $\alpha_c$. This behavior
is not shared by the standard Anderson model and has never been mentioned
before. Remarkably, the additional peak of the absorption is located close to
the low-energy mobility edge, thus providing a simple spectroscopic tool to
monitor it. We propose a simple explanation of the behavior found.

We consider a regular open chain of $N$ optically active two-level units with
parallel transition dipoles, which are coupled by the resonant dipole-dipole
interaction. The corresponding Hamiltonian reads
\begin{equation}
{\cal H}  = \sum_{n=1}^{N} \varepsilon_{n}\,|n\rangle\langle n| -
\sum_{n=1}^{N-1}\Big(|n\rangle\langle n+1|+|n+1\rangle\langle n|\Big)\ ,
\label{hamiltonian}
\end{equation}
where $|n\rangle$ denotes the state in which the $n$-th unit, having the energy
$\varepsilon_n$, is excited, whereas all the other units are in the ground
state. The intersite dipole-dipole coupling is restricted to nearest-neighbors
and set to $-1$ over the entire lattice.
$\varepsilon_1,\varepsilon_2,\ldots,\varepsilon_N$ is a stochastic long-range
correlated sequence. We generate a realization of the sequence according to the
following rule~\cite{Moura98}
\begin{equation}
    \varepsilon_{n} = \sigma \, C_\alpha \sum_{k=1}^{N/2}
    \frac{1}{k^{\alpha/2}}\,
    \cos\left(\frac{2\pi kn}{N}+\phi_k\right)\ .
    \label{disorder}
\end{equation}
Here, $C_\alpha = \sqrt{2}\big(\sum_{k=1}^{N/2} k^{-\alpha}\big)^{-1/2}$, and
$\phi_1,\phi_2,\ldots,\phi_{N/2}$ are $N/2$ independent random phases uniformly
distributed within the interval $[0,2\pi]$. The random
distribution~(\ref{disorder}) has zero mean $\langle \varepsilon_{n}\rangle = 0$
and a correlation function
\begin{equation}
    \langle \varepsilon_{n} \varepsilon_m \rangle
    =  \frac{\sigma^2C_\alpha^2}{2} \sum_{k=1}^{N/2} \frac{1}{k^\alpha}\,
    \cos\left[\frac{2\pi k(n-m)}{N}\right]\ ,
\label{correlator}
\end{equation}
where $\langle \ldots \rangle$ indicates an average over the distribution of
random phases $\phi_k$ and $N$ is assumed to be even. From~(\ref{correlator}) it
follows that $\sigma = \langle \varepsilon_{n}^2\rangle^{1/2}$ is the standard
deviation of the distribution~(\ref{disorder}). This quantity will be referred
to as magnitude of disorder. The long-range nature of the site potential
correlations results from the power-law dependence of the amplitudes of the
Fourier components in Eq.~(\ref{disorder}). As was shown in
Ref.~\onlinecite{Moura98}, a phase of extended states occurs at the band center
provided $\alpha\ > \alpha_c = 2$ when $\sigma = 1$.

The quantity of our primary interest is the absorption spectrum, which is
defined as
\begin{equation}
\label{A}
    A(E)= \frac{1}{N}\Bigg\langle \sum_{\nu=1}^{N}
    F_\nu \delta(E-E_{\nu})\Bigg\rangle \ ,
\end{equation}
where $E_\nu$ are the eigenenergies of the normalized eigenfunctions $\psi_{\nu
n}$ of the Hamiltonian~(\ref{hamiltonian}). The quantity $F_\nu =
\left(\sum_{n=1}^N \psi_{\nu n}\right)^2$ is the dimensionless oscillator
strength of the $\nu$-th state.

\begin{figure}[ht]
\includegraphics[width=0.82\columnwidth,clip]{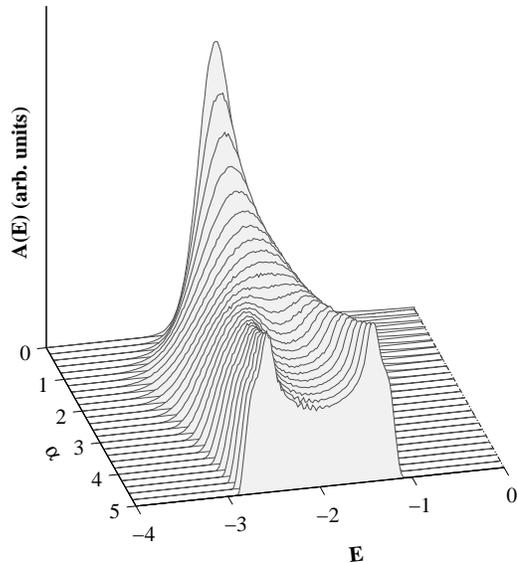}
\caption{Evolution of the absorption spectrum shape as a function of the power
exponent $\alpha$ and a given magnitude of disorder $\sigma = 1$. Notice the
appearance of a double-peaked structure of the absorption spectrum when $\alpha$
exceeds the critical value $\alpha_c = 2$ for the LDT to occur.} 
\label{fig1}
\end{figure}

We have numerically diagonalized the Hamiltonian~(\ref{hamiltonian}) and
obtained the absorption spectrum for different values of the power exponent
$\alpha$, considering open linear chains of size $N = 250$ and setting the
magnitude of disorder $\sigma = 1$. The results comprise an average over
$3\times 10^{4}$ realizations of the disorder for each value of $\alpha$.
Figure~\ref{fig1} shows the output of the simulations. When $\alpha \ll 1$ the
absorption spectrum displays a single and asymmetric peak slightly below the
lower band edge $E = -2$ of the periodic lattice ($\sigma=0$), i.e., only the
lowest states of the band contribute to the absorption spectrum. This trend is
exactly the same as that observed in 1D systems with uncorrelated randomness.
The only noticeable effect upon increasing $\alpha$ within the range $\alpha <
2$ is an increase of the absorption bandwidth. In other words, long-range
correlations in disorder cause a stronger localization as compared to the
uncorrelated case. A similar effect was found in 1D disorder-correlated systems
with a characteristic spatial length and explained on the basis of the exchange
narrowing concept~\cite{Malyshev99} (see also Ref.~\onlinecite{Russ98}).

The absorption shape changes dramatically when $\alpha > \alpha_c$, as seen in
Fig.~\ref{fig1}: A single peak splits into a doublet. One of the doublet
components (at low energy) is located at the bottom of the band as usual,
whereas the other one (at higher energy) lies deep inside the band. This means
that the oscillator strength does not decrease monotonously from the bottom to
the center of the band, indicating strong and unexpected effects of the
long-range correlated disorder on the spatial distribution of the probability
amplitude. In contrast to the case $\alpha < 2$, the broadening of the peaks
drops down on increasing $\alpha$ and then saturates. Finally, notice that the
low-energy tail of the low-energy peak loses its characteristic Gaussian shape
observed in uncorrelated disordered systems. All these features point out a rich
phenomenology of the model under study that cannot be accounted for within the
standard theoretical frameworks.

\begin{figure}[ht]
\includegraphics[width=0.82\columnwidth,clip]{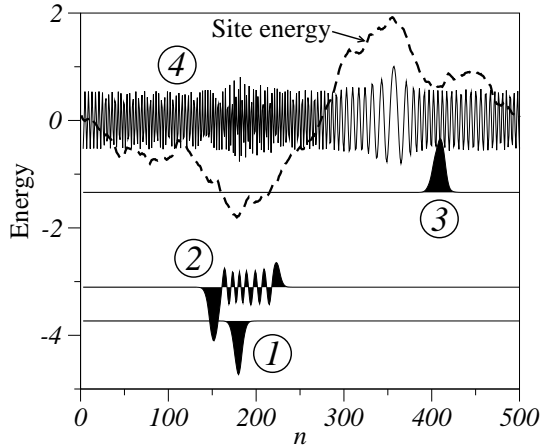}
\caption{A subset of eigenstates for a typical realization of the random energy
potential (dashed line), for a chain of size $N=500$, magnitude of disorder
$\sigma = 1$, and power exponent $\alpha=3.0$ (larger than the critical value
$\alpha_c = 2$). The baselines indicate the energies of each eigenstate. The
states 1 and 3 are those which contribute to the low- and high-energy peaks of
the absorption spectrum, respectively.}
\label{fig2}
\end{figure}

To get insight into the effects of scale-free disorder on the optical properties
of the system, it is useful to have a look at the eigenfunctions of the
Hamiltonian~(\ref{hamiltonian}). In Fig.~\ref{fig2} we depict a subset of wave
functions obtained for a typical random realization of the potential landscape.
The baselines display the corresponding eigenenergies. The lowest state in
Fig.~\ref{fig2} (labelled by $1$) is localized in the sense that it has a
considerable amplitude within a domain of size $N^*$, which is smaller than the
system size $N$. It shows a bell-like shape and carries a large oscillator
strength $F_1 \propto N^*$. There are several states of such type (not shown in
Fig.~\ref{fig2}), which are close in energy to the state $1$ and do not overlap
with each other. They contribute to the low-energy peak of the absorption
spectrum. On increasing the energy, one observes eigenstates, like the one
labelled by $2$ in Figure~\ref{fig2}, which are more extended, as compared to
the lowest one, and present several nodes within the localization segment. The
oscillator strength of such states is smaller than $F_1$. Consequently, they
contribute to the absorption spectrum to a lesser extent. Remarkably, going
further up in energy we again find bell-like states, as the one labelled by $3$
in Figure~\ref{fig2}, which are characterized by a large oscillator strength.
Those states form the high-energy peak of the doublet in the absorption
spectrum. Finally, on approaching the center of the band one expects the
occurrence of extended states. The state $4$ in Figure~\ref{fig2} represents an
example.

\begin{figure}[ht]
\includegraphics[width=0.82\columnwidth,clip]{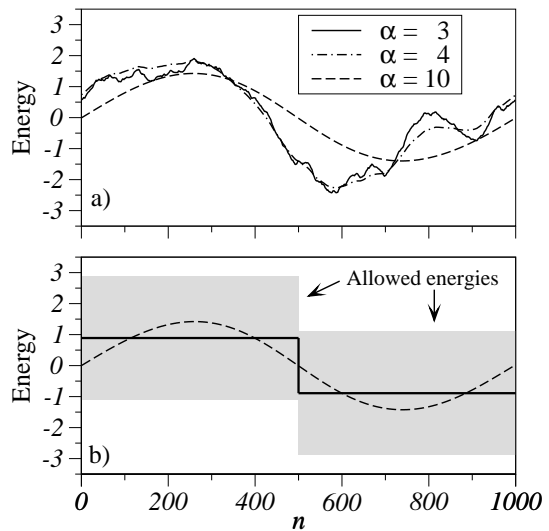}
\caption{(a) The site energy landscape $\varepsilon_n$ given
by~(\protect{\ref{disorder}}) for $\sigma=1$ and three different values of
$\alpha$. The random phases have been shifted the same amount so that
$\phi_1=-\pi/2$. (b) A model in which the actual site energy landscape (dashed
line) is replaced by a step-like energy profile (solid line). The shaded regions
indicate the allowed energy bands of each segment, whose width is $4$ in the
chosen units.} 
\label{fig3}
\end{figure}

Aiming to elucidate the anomalies found, we present a simplified model that
explains the occurrence of the optically active states deep inside the band as
well as sheds light on the nature of the LDT in the model under study. To this
end, we recall that the site energy potential~(\ref{disorder}) is given by a sum
of spatial harmonics. The amplitude of each term, $\sigma C_\alpha
k^{-\alpha/2}$, decreases upon increasing the harmonic number $k$. For
sufficiently high values of $\alpha$, the first term in the
series~(\ref{disorder}) will be dominant, while the others are considerably
smaller. Consequently, the site potential for a given realization represents a
deterministic function of period $N$ (harmonic with $k=1$), perturbed by a
colored noise (harmonics with $k\geq 2$). Figure~\ref{fig3}(a) shows the site
energy landscape~(\ref{disorder}) for different values of $\alpha > 2$,
illustrating the statement above. Therefore, relevant information can be
obtained by analyzing the first ($k = 1$) term in the series~(\ref{disorder}).

As a further step of simplification, we replace the sine-like site energy
potential by a step-like one, as shown in Fig.~\ref{fig3}(b). More specifically,
we take $\varepsilon_n = \bar{\varepsilon}\,\text{sign} (N/2 - n)$, where
$\bar{\varepsilon} = (2/\pi)\sigma C_\alpha = 0.63\sigma C(\alpha)$ is the
average value of the site energies on the left half of the system. In doing so,
we map the original lattice onto two uniform sublattices, coupled to each other
through the hoping between sites $N/2$ and $N/2+1$. The allowed energies of each
\emph{decoupled\/} sublattice form a band (referred hereafter, for the sake of
clarity, to as subband), ranging from $\bar{\varepsilon} - 2$ to
$\bar{\varepsilon} + 2$ and from $-\bar{\varepsilon} - 2$ to $-\bar{\varepsilon}
+ 2$ for the left and right sublattices, respectively [see Fig.~\ref{fig3}(b)].
The absorption spectrum of such a system is expected to have two peaks caused by
the transitions from the ground state to the bottom state of each subband. For
$\sigma = 1$ their locations are $\bar{\varepsilon} - 2 = -1.11$ and
$-\bar{\varepsilon} - 2 = -2.89$. We stress that these values are in a fairly
good agreement with the results of exact calculations presented in
Fig.~\ref{fig1}, despite both the simplification involved and neglecting the
coupling. This is a strong indication that the lowest states of the subbands are
not very much modified after switching on the interaction between sublattices.

In order to understand the above features, let us consider the elements of the
coupling matrix between segments, given by~\cite{Malyshev99}
\begin{equation}
\label{V}
    V_{k_l k_r} = (-1)^{k_l} \frac{4}{N + 2}\,
    \sin\left(\frac{2\pi k_l}{N + 2}\right)\,
    \sin\left(\frac{2\pi k_r}{N + 2}\right) \ .
\end{equation}
Here, $k_l$ and $k_r$, ranging from 1 to $N/2$, number the eigenstates of the
left and right sublattices, respectively. From~(\ref{V}) we find that the
magnitude of coupling of the lowest state of the left sublattice ($k_r = 1$) to
the closest state of the left one ($k_l = 1$) is $|V_{11}| \approx 16\pi^2/(N +
2)^3$, that is much smaller than $2\bar{\varepsilon}=1.78$, the energy
difference between the lowest band edges (the limit of $N \gg 1$ is implied).
The magnitude of the coupling of the lowest state of the left sublattice ($k_l =
1$) to the central band states of the right one is about $|V_{1k_r}| \approx
8\pi/(N + 2)^2$, whereas the energy spacing at the center of the band is
$4\pi/(N + 2)$. Again, the coupling is smaller than the energy spacing. This
explains the low sensitivity of the lowest states of the subbands to switching
on the interaction between the sublattices.

\begin{figure}[ht]
\includegraphics[width=0.82\columnwidth,clip]{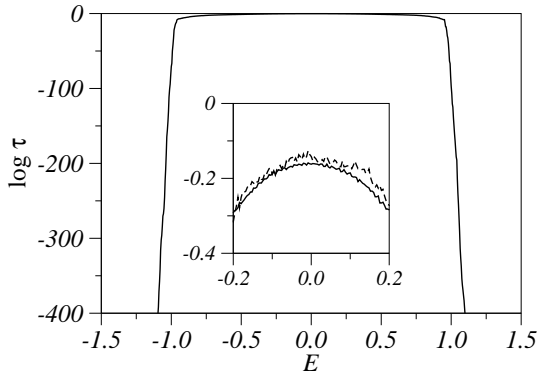}
\caption{Log-normal plot of the transmission coefficient $\tau$ as a function of
the incoming energy when $\alpha=4.0$ and $N=3\times 10^3$. The inset represents
an enlarged view of $\tau$ close to the center of the band for $N=3\times 10^3$
(solid line) and $N=3\times 10^4$ (dashed line), showing that $\tau$ does not
change on increasing the size. $3\times 10^3$ disorder realizations were
considered.} 
\label{fig4}
\end{figure}

Note the existence of an overlap of the two subbands shown in
Fig.~\ref{fig3}(b). The overlap region provides a pass-band filter for a
quasiparticle. This is a crucial ingredient in understanding the origin of the
appearance of both a phase of extended states at the center of the band and the
LDT in the presence of disorder. Following the same reasoning as before, we can
conclude that upon increasing $\alpha$, the role of the random phases
$\phi_2,\ldots,\phi_{N/2}$ in the series~(\ref{disorder}) gets smaller and
smaller as compared to the leading term $k = 1$. Consequently, one ends up with
our simplified model to explain the phase of extended states for higher values
of $\alpha$.

The cutoff energies of the pass-band can be associated with the mobility edges.
The actual locations of the latter can be estimated by studying the
transmission. In Fig.~\ref{fig4} we plotted the transmission coefficient $\tau$
as a function of the incoming energy, which was calculated for $\sigma = 1$ and
$\alpha = 4 > \alpha_c = 2$ (about the procedure for calculating $\tau$ see,
e.g. Ref.~\onlinecite{Macia00}). The presence of a pass-band is clearly seen.
Furthermore, we observe that the low-energy transmission cutoff, $E \approx -1$,
is very close to the location of the high-energy peak in the absorption
spectrum, $\bar{\varepsilon} - 2 = -1.11$. This provides the link between the
second peak in the doublet of the absorption spectrum and the low-energy
mobility edge.

In summary, we have studied numerically the linear optical response of a
quasiparticle moving in the long-range correlated energy landscape with a
power-like spectrum $S(k)\sim 1/k^{\alpha}$. We found a crossover of the
absorption lineshape from a single-peaked form to a doublet when varying the
power exponent $\alpha$ from zero to $\alpha > \alpha_c$, where $\alpha_c$ is
the critical value for a LDT to occur in the present model; this signals the
occurrence of the localization-delocalization transition. The low-energy peak is
located at the bottom of the band, whereas the high-energy peak lies deep inside
of the band, indicating the presence of band states with a large oscillator
strength. The location of the high-energy peak is slightly below the low-energy
mobility edge of the phase of extended states. This provides a unique
possibility to monitor the mobility edge spectroscopically. The crossover found
is a characteristic feature of the underlying long-range correlated random
sequence and, thus, can also be used to distinguish it from other types of
correlations.

\acknowledgments

The authors thank A.\ V.\ Malyshev, M.\ L.\ Lyra and F.\ A.\ B.\ F.\ de Moura
for helpful comments. This work was supported by DGI-MCyT (Project
MAT2003-01533) and CAM (Project GR/MAT/0039/2004). V.\ A.\ M.\ acknowledges
support from ISTS (grant \#2679).


\begin{thebibliography}{99}

\bibitem{Anderson58} P.\ W.\ Anderson, Phys.\ Rev.\ \textbf{109},
    1492 (1958).

\bibitem{Abrahams79} E.\ Abrahams, P.\ W.\ Anderson, D.\ C.\
    Licciardello, and  T.\ V.\ Ramakrishnan, Phys.\ Rev.\ Lett.\
    \textbf{42}, 673 (1979).

\bibitem{Beenakker97} C.\ W.\ J.\ Beenakker, Rev.\ Mod.\ Phys.\
    \textbf{69}, 731 (1997).

\bibitem{Janssen98} M.\ Janssen, Phys.\ Rep.\ \textbf{295}, 1
    (1998).

\bibitem{Flores89} J.\ C.\ Flores, J.\ Phys.:\ Condens.\ Matter
    \textbf{1}, 8471 (1989).

\bibitem{Phillips91} P.\ Phillips and H.-L.\ Wu, Science
    \textbf{252}, 1805 (1991).

\bibitem{Flores93} J.\ C.\ Flores and M.\ Hilke, J.\ Phys.\ A
    \textbf{26}, L1255 (1993).

\bibitem{Adame93} F.\ Dom\'{\i}nguez-Adame, E.\ Maci\'{a}, and
    A.\ S\'{a}nchez, Phys.\ Rev.\ B \textbf{48}, 6054 (1993).

\bibitem{Bellani99} V.\ Bellani, E.\ Diez, R.\ Hey, L.\ Toni, L.\
    Tarricone, G.\ B.\ Parravicini, F.\ Dom\'{\i}nguez-Adame, and
    R.\ G\'{o}mez-Alcal\'{a}, Phys.\ Rev.\ Lett.\ \textbf{82},
    2159 (1999).

\bibitem{Moura98} F.\ A.\ B.\ F.\ de Moura and M.\ L.\ Lyra, Phys.\
    Rev.\ Lett.\ \textbf{81}, 3735 (1998).

\bibitem{Kuhl00} U.\ Kuhl, F.\ M.\ Izrailev, A.\ A.\ Krokhin, and
    H.\ -J.\ St\"{o}ckmann, Appl.\ Phys.\ Lett.\ \textbf{77}, 633
    (2000).

\bibitem{Izrailev99} F.\ M.\ Izrailev and A.\ A.\ Krokhin, Phys.\
    Rev.\ Lett.\ \textbf{82}, 4062 (1999).

\bibitem{Carpena02} P.\ Carpena, P.\ B.\ Galv\'{a}n, P.\ Ch.\ Ivanov,
    and H.\ E.\ Stanley, Nature {\bf 418}, 955 (2002); {\bf 421},
    764 (2003).

\bibitem{Yamada04} H.\ Yamada, Phys.\ Lett.\ A \textbf{332}, 65
    (2004); Int. J. Mod. Phys. B \textbf{18}, 1697 (2004).

\bibitem{Carpena04} P.\ Carpena, P.\ Bernaola-Galv\'{a}n, and P.\ Ch.\
    Ivanov, Phys.\ Rev.\ Lett.\ \textbf{93}, 176804 (2004).

\bibitem{Adame03} F.\ Dom\'{\i}nguez-Adame, V.\ A.\ Malyshev,
    F.\ A.\ B.\ F.\ de Moura, and M.\ L.\ Lyra, Phys.\ Rev.\ Lett.\
    \textbf{91}, 197402\ (2003).

\bibitem{Malyshev99} V.\ A.\ Malyshev, A.\ Rodr\'{\i}guez, and F.\
    Dom\'{\i}nguez-Adame, Phys.\ Rev.\ B \textbf{60}, 14\,140 (1999).

\bibitem{Russ98} S.\ Russ, S. Havlin, and I. Webman, Phil. Mag. B
    {\bf 77}, 1449 (1998).

\bibitem{Macia00} E.\ Maci\'{a} and F.\ Dom\'{\i}nguez-Adame,
    \emph{Electrons, phonons and excitons in low-dimensional
    aperiodic systems} (Editorial Complutense, Madrid, 2000).

\end{thebibliography}
\end{document}